\documentclass[aps,prd,preprint,superscriptaddress,tightenlines,nofootinbib]{revtex4}

\usepackage{graphicx}
\usepackage{dcolumn}
\usepackage{bm}

\usepackage{epsfig}

\newcommand{\etal}{{\it et al.}}

\begin{document}

\preprint{CLNS 02/1805}       
\preprint{CLEO 02-14}         

\title{Inclusive $\eta'$ Production From The $\Upsilon (1S)$}
\author{M.~Artuso}
\author{C.~Boulahouache}
\author{S.~Blusk}
\author{K.~Bukin}
\author{E.~Dambasuren}
\author{R.~Mountain}
\author{H.~Muramatsu}
\author{R.~Nandakumar}
\author{T.~Skwarnicki}
\author{S.~Stone}
\author{J.C.~Wang}
\affiliation{Syracuse University, Syracuse, New York 13244}
% ...
\collaboration{CLEO Collaboration} 
\noaffiliation
\author{(Dated: November 11, 2002)}\noaffiliation

\begin{abstract}
Using the CLEO II detector at CESR, we measure the $\eta'$ energy spectra
in $\Upsilon(1S)$ decays, that we compare with models of the $\eta'g^*g$
form-factor. This form-factor especially
at large $\eta'$ energies may provide an explanation of the large
rate for $B\to X_s\eta'$. Our data do not support a large
anomalous coupling at higher $q^2$ and thus the large $\eta'$ rate
remains a mystery, possibly requiring a non-Standard Model
explanation.
\end{abstract}

\pacs{13.25.Gv, 13.25.Hw, 13.25.-k, 13.65.+i}

\maketitle
\section{Introduction}
There are several interesting, unexplained phenomena in $B$
decays. First of all, the total production of charm and charmonium
seems about 10\% low \cite{charm_deficit}, especially when coupled with a
$B$ semileptonic branching ratio of (10.4$\pm$0.3)\% \cite{PDG}. Secondly, CLEO observed
a very large rate of $\eta'$ in the momentum range from 2 to 2.7
GeV/c with a branching fraction of $(6.2\pm 1.6\pm
1.3^{+0.0}_{-1.5})\times 10^{-4}$ \cite{Browder}. The BABAR experiment
has confirmed this large rate \cite{Babaretap}. The production of
$\eta'$ mesons is believed to occur dominantly via the $b\to s g$ 
mechanism, as strongly
suggested by observation of the two-body decay $B\to \eta' K$. One
explanation of the large $\eta'$ rate is that the $b\to s g$ rate
is not 1\% as expected in the Standard Model, but is enhanced
by new physics to be at the 10\% level. This would also explain
the charm deficit problem.

\begin{figure}[htb]
  \begin{center}
    \leavevmode
    \epsfxsize=80mm
    \epsfbox{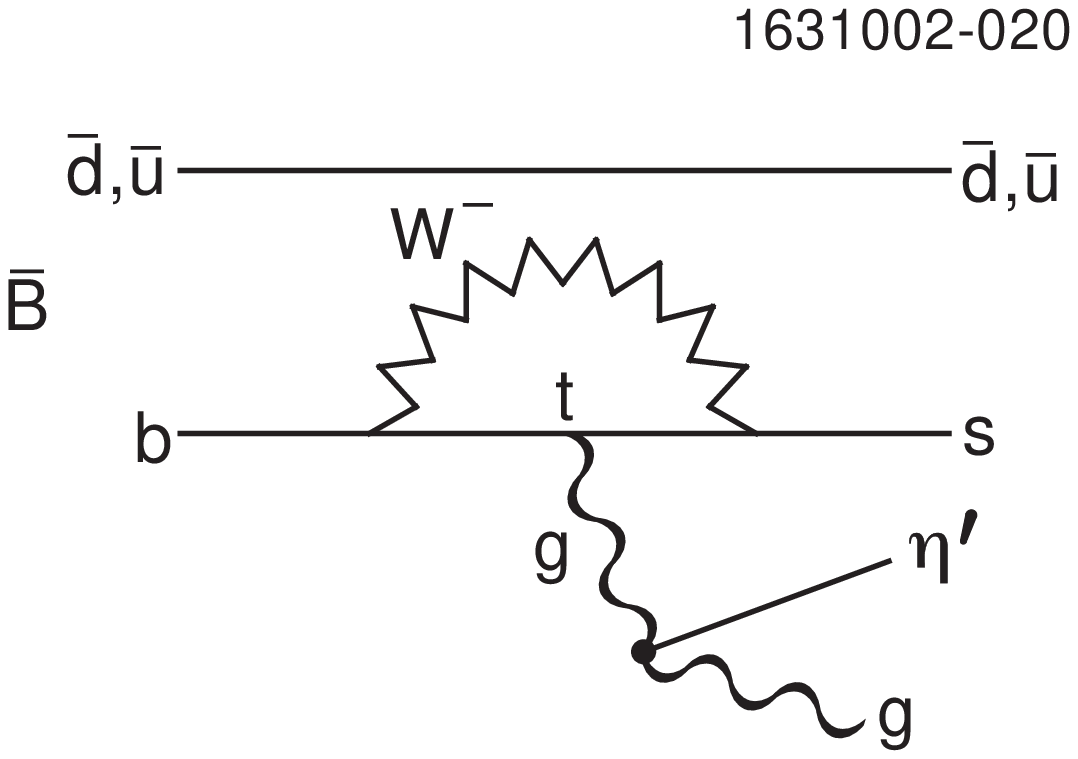}
    \caption{Diagram for $b\to sg\eta'$.
    }
    \label{fig:btoetap}
  \end{center}
\end{figure}

An alternative explanation is that of an anomalously strong
coupling between the $\eta'$ and two gluons \cite{Atwood:1997bn,Hou:1997wy,Kagan:1997sg}.
The process $b\to s g$ followed by the two gluon coupling to the
$\eta'$ is shown in Fig.~\ref{fig:btoetap}.

Experimentally, the hadronic mass associated with $X_s$ sometimes
is a $K$, $\sim$10\%, and even more rarely a $K^*$, $\sim$1\%; in fact,
most of the rate has the mass of the $X_s$ system larger than 1.8
GeV.

Since the $\eta'$ is mostly the flavor singlet $\eta_1$, as the
$\eta-\eta'$ mixing angle is between $10^{\circ}-20^{\circ}$, the
effective $\eta'g^*g$ coupling can be written as \cite{Kagan02}
\begin{equation}
H(q^2)\varepsilon_{\alpha\beta\mu\nu}q^{\alpha}k^{\beta}\varepsilon^{\mu}_{1}
\varepsilon^{\nu}_{2},
\end{equation}
where $q=p_b-p_s$ is the four-momentum of the virtual hard gluon $(g^*)$, $k$ is the four-momentum of the soft ``on-shell" gluon $(g)$, and $H(q^2)$ is the $g^*g\eta'$ transition form factor.

Chen and Kagan \cite{Kagan02} have shown that the
region of the $q^2$ relevant in the process $b \to sg\eta'$ can
also be accessed in high energy $\eta'$ production in $\Upsilon
(1S)$ decay. Thus constraints can be put on the $H(q^2)$ from the
$\eta'$ spectrum in $\Upsilon (1S)\to ggg$ decays. $H(0)$ is found from
the rate of $J/\psi\to \gamma\eta'$ decays as $\sim$1.8
GeV$^{-1}$.

\begin{figure}[htb]
  \begin{center}
    \leavevmode
    \epsfxsize=100mm
    \epsfbox{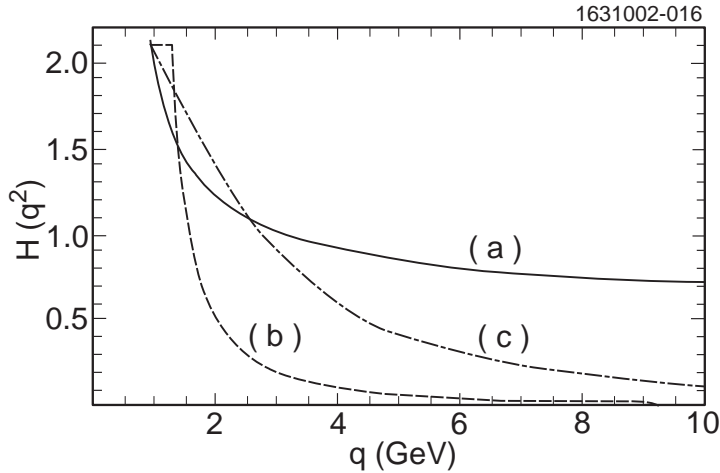}
    \caption[A figure imported from a PostScript file]{
Three choices for the form factor $H(q^2)$ plotted against $\sqrt{q^2}$: (a) the
slowly falling form factor, (b) a rapidly falling form factor
representative of perturbative QCD calculations, (c) an
intermediate example (adapted from \cite{Kagan02}).}
    \label{fig:q2dep}
  \end{center}
\end{figure}

Three choices for the form factor shape $H(q^2)$ are shown in
Fig.~\ref{fig:q2dep}: (a) a slowly falling form factor from Hou
and Tseng \cite{Hou:1997wy}, $H(q^2)=2.1~{\rm
GeV}^{-1}\alpha_s(q^2)/\alpha_s(m^2_{\eta'})$; (b) a rapidly
falling form factor representative of perturbative QCD
calculations, $H(q^2)=1.7$
GeV$^{-1}$m$^2_{\eta'}/(q^2-m^2_{\eta'})$ at $q^2>$ 1 GeV$^2$; (c)
an intermediate example with $H(q^2)\propto 1/(q^2+2.2^2{\rm
GeV}^2)$ \cite{Kagan02}. In (b) and (c) the form factor at $q^2
\approx m_{\eta{\prime}}^2 $ has
been matched onto the value
given in (a), which is fixed by the QCD anomaly \cite{Hou:1997wy}.
The parametrization of the form factor in (b) follows from a simple model
in which the
$\eta^\prime $ is coupled perturbatively to two gluons through quark loops
\cite{Kagan:1997sg}.
With the choice $H(0) = 1.7$ GeV$^{-1}$ it compares
well with the perturbative QCD form factors obtained by other authors
\cite{Ali:2000ci,Muta}.

We will compare the theoretical predictions for $H(q^2)$ with data
taken on the $\Upsilon (1S)$ resonance with the CLEO II detector
at the CESR storage ring. Some information on this topic has been
extracted by Kagan from ARGUS data \cite{Kagan_argus}.

\section{Data Sample and Analysis Method}
In this study we use 80 pb$^{-1}$ of CLEO II data recorded at the $\Upsilon (1S)$ resonance (9.46 GeV), containing $1.862 \times 10^6$
$\Upsilon (1S)$ events. We also use off-resonance continuum data
collected below the $\Upsilon(4S)$ resonance (10.52 GeV) with a total integrated
luminosity of 1193 pb$^{-1}$.

The theoretical predictions referred to in this paper are made for
$\Upsilon (1S)$ decays into three gluons ($ggg$). In order to
compare our measurement to them we have to correct for the
$\Upsilon (1S)\to\gamma^*\to q\bar{q}$ contribution, whose size is
given by
\begin{equation}
{\cal{B}}(\Upsilon (1S)\to\gamma^*\to q\bar{q})=R\cdot
{\cal{B}}(\Upsilon (1S)\to\mu^+\mu^-)=(8.8\pm0.3)\%,
\end{equation}
where $R_{\sqrt{s}\approx9.5}=3.56\pm0.07$ \cite{Ammar:1997sk} and
the ${\cal{B}}(\Upsilon (1S)\to\mu^+\mu^-)$ is taken as
$(2.48\pm0.06)\%$ \cite{PDG}.

Although several processes can contribute to inclusive $\eta'$
production in $\Upsilon (1S)$ decays, it is believed that the soft
processes including fragmentation populate only the low $q^2$ or
equivalently the low Z region, where
\begin{equation}
{\rm Z}\equiv E_{\eta'}/E_{beam}=2E_{\eta'}/M(\Upsilon (1S)).
\end{equation}
Thus in the large Z region significant
$\eta'$ production would indicate a large $\eta'g^* g$ coupling.

The CLEO II detector, described in detail elsewhere \cite{CLEO_II}
had a high resolution electromagnetic calorimeter comprised of
7800 CsI crystals surrounding a precision tracking system.

We detect $\eta'$ mesons using the decay channel: $\eta' \to
\eta\pi^+\pi^-$ with a branching fraction of $44\%$, and $\eta \to
\gamma\gamma$ with a branching fraction of $39\%$. We identify
single photons based on their shower shape and the non-proximity
of charged tracks. Those photon pairs within the ``good barrel"
region of the detector, $|\cos\theta| <0.707$ (where $\theta$ is
the angle with respect to the beam), that have invariant masses
consistent with the $\eta$ mass within 3 standard deviations are
constrained to have the invariant mass of the $\eta$. For $\eta$
mesons coming from low energy $\eta'$ candidates ($\rm Z < 0.5 $)
the background from $\pi^\circ$ decay is large, and thus the candidate
photons are also required not to be from a possible
$\pi^\circ$ decay. We then add two opposite sign pions and form
the $\eta\pi^+\pi^-$ invariant mass.

\begin{figure}[htb]
\vspace{-4mm}
  \centerline{\epsfig{figure=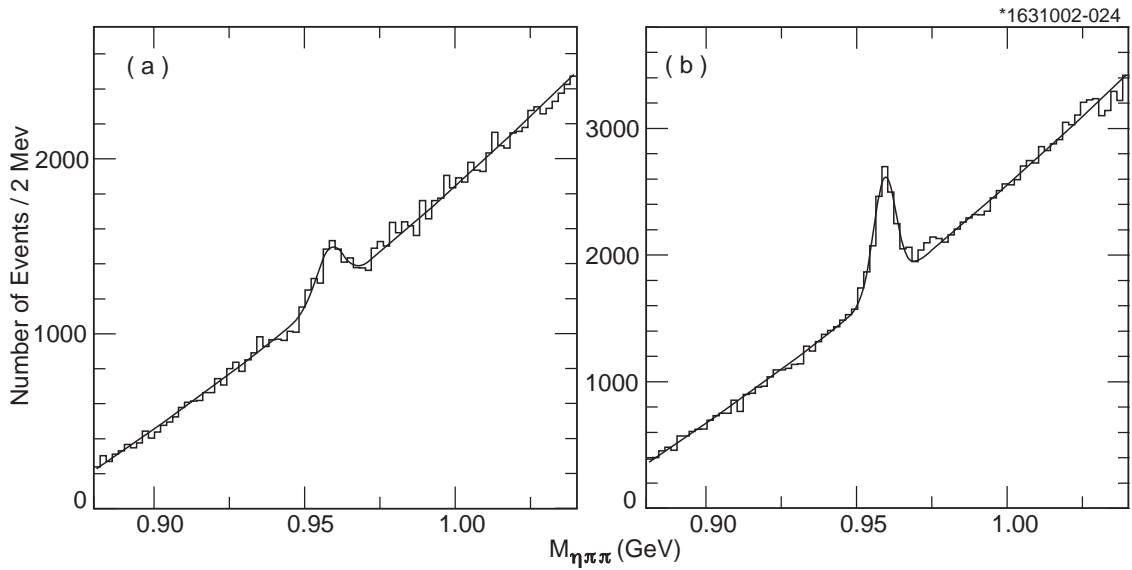,height=3in}}
\vspace{-4mm}
  \caption{\label{fig:xm1sa} The $\eta\pi^+\pi^-$ invariant mass spectrum
   reconstructed from $\Upsilon (1S)$ data (left), and off resonance
   data at 10.52 GeV (right) fit with Gaussian functions for signal
   and second order polynomials for background.}
\end{figure}

The $\eta\pi^+\pi^-$ invariant mass spectra are shown in
Fig.~\ref{fig:xm1sa} for $\Upsilon (1S)$ and for
off-resonance continuum data. The spectra are fit with a Gaussian
function for signal and second order polynomial function for
background. The numbers of reconstructed $\eta'$ are extracted
from the fit. We find $\rm 1486 \pm 137\ \eta'$ from the $\Upsilon
(1S)$ data, and $\rm 4062 \pm 174\ \eta'$ from the off-resonance
data.

\begin{figure}[htbp]
  \centerline{\epsfig{figure=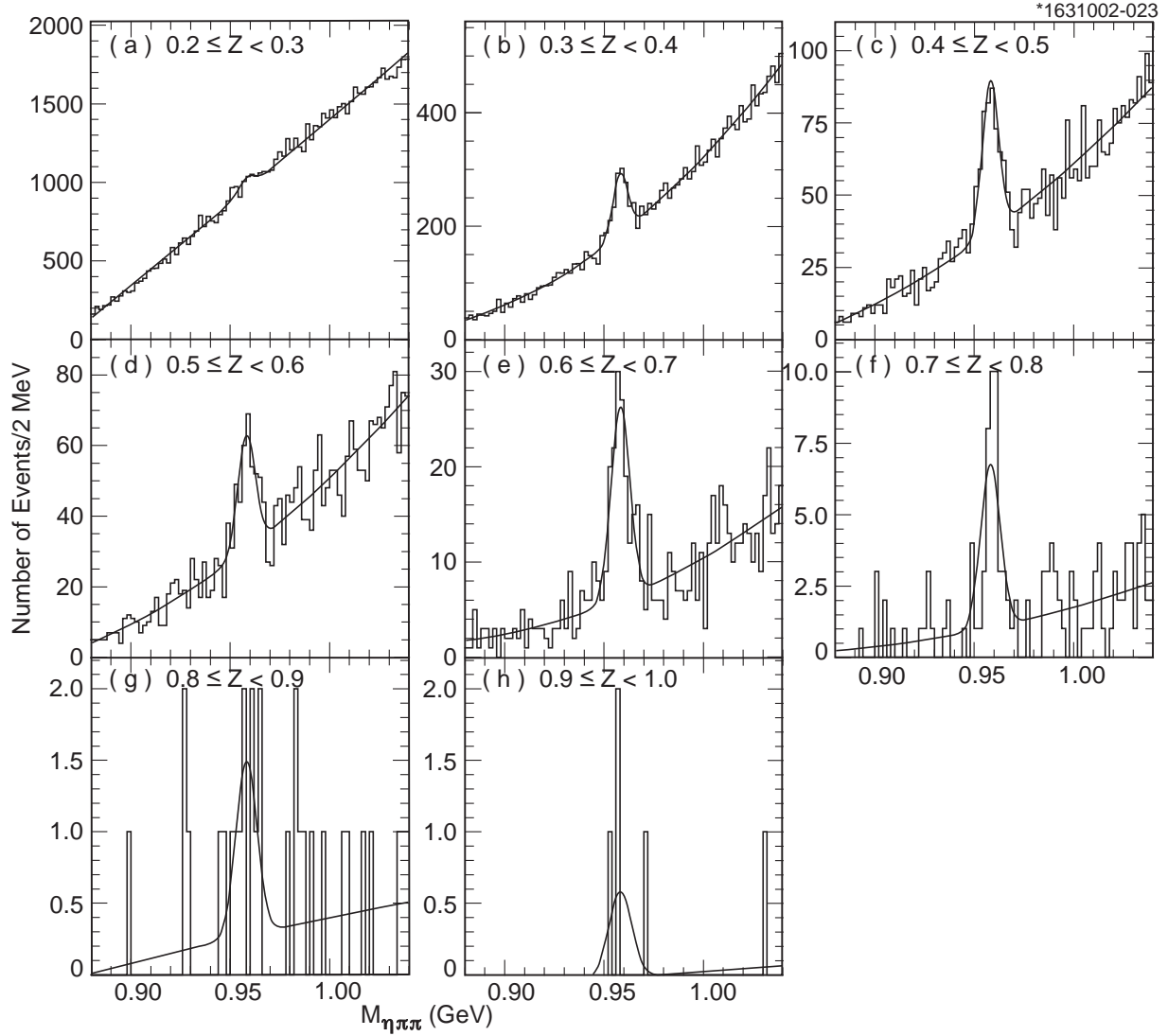,height=5.8in}}
\vspace{-1mm}
  \caption{\label{fig:xm1sz} The $\eta\pi^+\pi^-$ invariant mass spectra in
  different Z ranges reconstructed from $\Upsilon (1S)$ data,
  fit with a Gaussian function for signal and a second order polynomial for background.}
\end{figure}

\begin{figure}[htbp]
  \centerline{\epsfig{figure=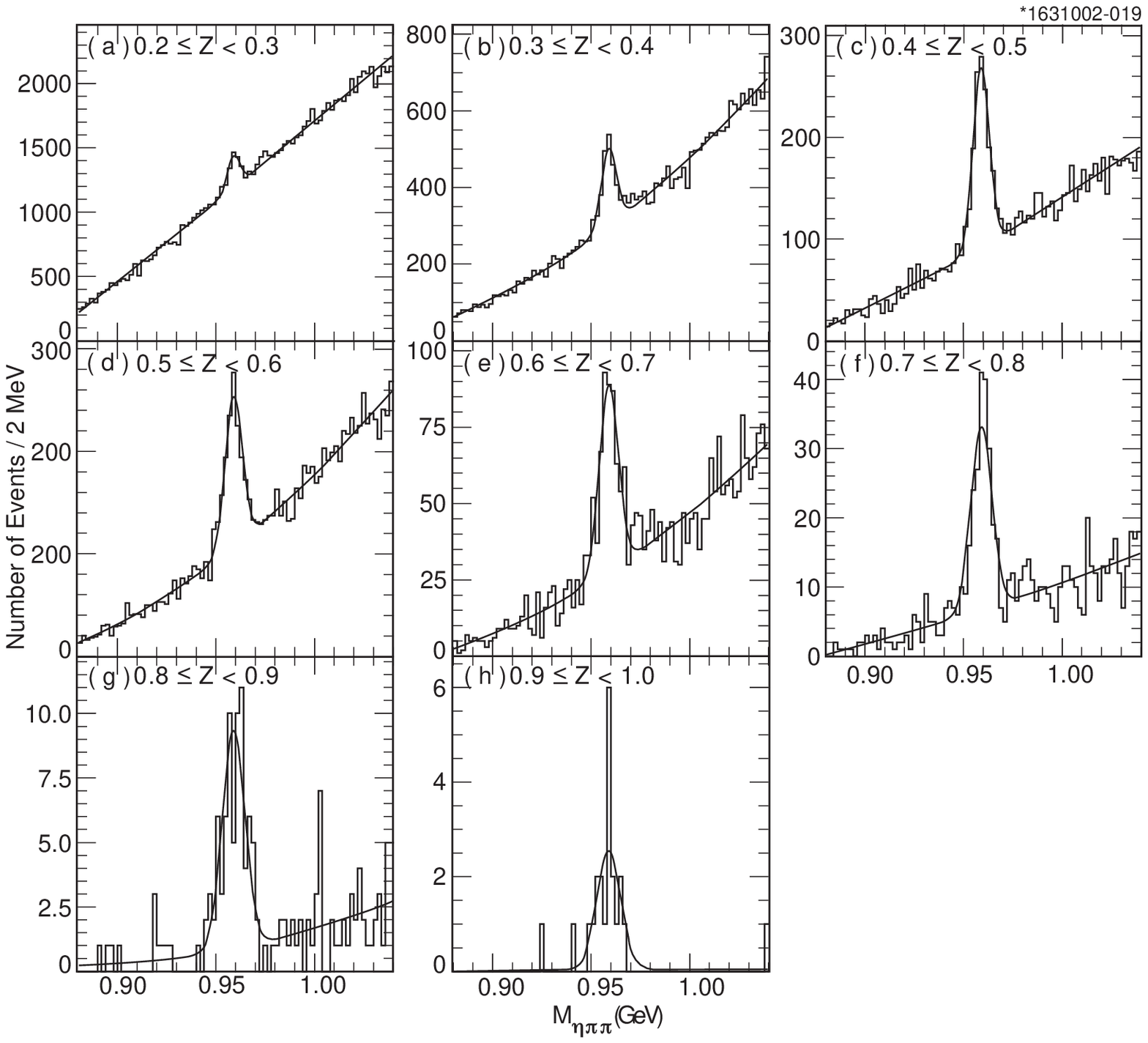,height=6.2in}}
\vspace{-16mm}
  \caption{\label{fig:xm4sz} The $\eta\pi^+\pi^-$ invariant mass spectra in
  different Z ranges reconstructed from off-resonance data,
  fit with a Gaussian function for signal and a second order polynomial for background.}
\end{figure}

To measure the energy spectrum we reconstruct $\eta'$ candidates
in Z intervals. We choose the Z steps as 0.1. The invariant mass
spectra are fit with the same functional form as used for
Fig.~\ref{fig:xm1sa}. Here we fix the mass of the $\eta'$ to our
average value over all Z; the Monte Carlo simulation shows that the mass
measurement should be independent of $\eta'$ energy. We extract
the width of the signal Gaussian distribution from Monte Carlo
simulation for each Z bin and perform a smooth fit as a function
of Z. The smoothed values are used in the fit as fixed
parameters. The $\eta\pi^+\pi^-$ Z dependent mass spectra are
shown in Fig.~\ref{fig:xm1sz} and Fig.~\ref{fig:xm4sz} for
$\Upsilon (1S)$ and off-resonance data, respectively.



In order to extract decay rates we need to correct our raw event
yields by efficiencies. These may not be equal for different intermediate
states, i.e. $q\bar{q}$ versus $ggg$. The hadronic events at
$\Upsilon (1S)$ energy arise from different sources: about 4 nb is
$q\bar{q}$ from continuum $\rm e^+e^-$ collisions, about 2 nb from
$\Upsilon (1S)\to\gamma^*\to q\bar{q}$, 18 nb from $ggg$, and 0.5
nb from $\gamma gg$ from the $\Upsilon (1S)$. The first two have
same event topology and reconstruction efficiencies. We use the
$q\bar{q}$ Monte Carlo generator to simulate these events. The
$\gamma gg$ events are similar to that of $ggg$ and have a
relatively small cross-section; thus we treat them the same way as
$ggg$ events. We use the $ggg$ Monte Carlo generator to simulate
this part.

We rely on off-resonance continuum data to estimate the $q\bar{q}$
contribution in $\Upsilon (1S)$ data. However, the continuum data
were taken for continuum subtraction in $\Upsilon(4S)$ studies. The center
of mass (CM) energy (10.52 GeV) is close to $\Upsilon(4S)$
mass (10.58 GeV), but more than 1 GeV higher than $\Upsilon (1S)$
mass (9.46 GeV). The difference of reconstruction efficiency due
to this energy difference is not negligible. We thus use different
$q\bar{q}$ simulations for continuum data and $\Upsilon (1S)$
data.

\begin{figure}[htb]
  \centerline{\epsfig{figure=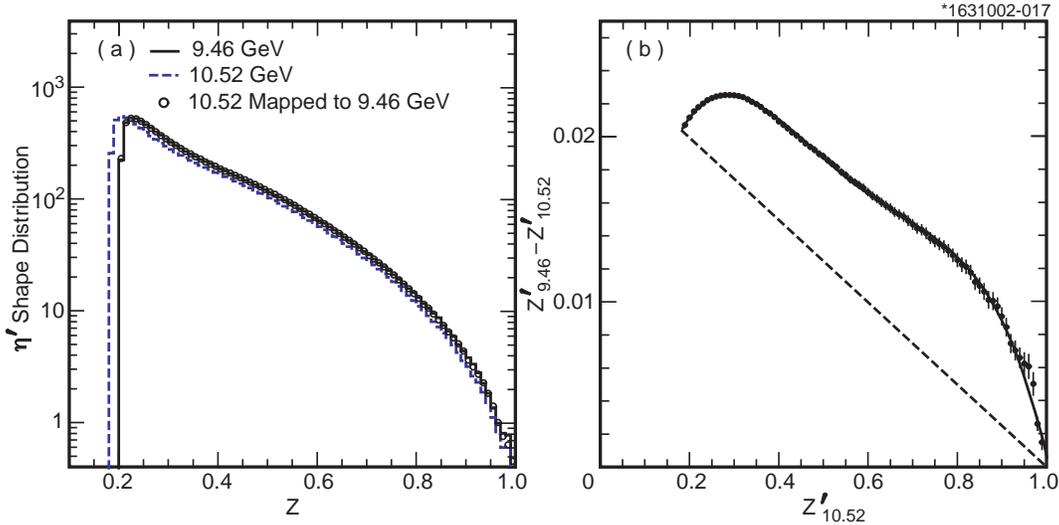,height=2.8in}}
\caption{\label{fig:znew} a) The Z $=E_{\eta'}/E_{beam}$ distributions
from Monte Carlo
simulation. The solid line is the Z=$E_{\eta'}/E_{beam}$ spectrum for an energy
of 9.46 GeV,
the dashed line is the spectrum for 10.52 GeV and the open circles are the
mapped spectrum from 10.52 GeV. b) The data points show the difference in
the Z values at 9.46 and 10.52 GeV as a function of the Z value at 10.52 GeV.
The solid curve is a fit to a fourth order polynomial. The dotted line shows
the mapping of the linear conversion.}
\end{figure}

The energy difference also affects the Z spectrum of $\eta'$ from
continuum Monte Carlo as shown in Fig.~\ref{fig:znew}-(a). The solid line
is the $\rm E_{\eta'}/E_{beam}$ distribution for the $\Upsilon
(1S)$ data (9.46 GeV) and dashed line for the
continuum data (10.52 GeV). The low limits are
0.202 and 0.182 respectively. The discrepancy is significant,
especially at low energy. In order to use our continuum data at
10.52 GeV we need to map it to 9.46 GeV. To do so we rely on the
continuum Monte Carlo. We take the two Monte Carlo $\eta'$ shape
distributions at 10.52 and 9.46 GeV, denoted by ${\cal
P}_{10.52}(z)$ and ${\cal P}_{9.46}(z)$ and numerically integrate
them to satisfy the relation:
\begin{equation}
\int^{{\rm Z}'_{10.52}}_0 {\cal P}_{10.52}(z) dz =
\int^{{\rm Z}'_{9.46}}_0 {\cal P}_{9.46}(z) dz~~,
\end{equation}
where Z$'_{10.52}$ is fixed and a value for Z$'_{9.46}$ is
determined. The data points on Fig.~\ref{fig:znew}-(b) show the
difference in Z$'_{9.46} - $Z$'_{10.52}$ as a function of
Z$'_{10.52}$ (or equivalently $\rm Z_0$ in following function). We
fit the points with a fourth order polynomial function to define
the mapping analytically as
\begin{equation}\label{eq:Z}
\rm Z = -0.215 \times 10^{-2} + 1.2238\ Z_0 - 0.6879\ Z_0^2
               + 0.8277\ Z_0^3 - 0.3606\ Z_0^4\ .
\end{equation}
The simplest mapping would be a linear conversion $Z = 0.025 +
0.975 \times Z_0$, shown as dotted line in
Fig.~\ref{fig:znew}-(b). We use this alternative to estimate the
systematic uncertainty due to the mapping.

That this mapping works is demonstrated in
Fig.~\ref{fig:znew}-(a), where the spectra shown as open circles
is the mapped spectrum according to Equation~\ref{eq:Z}. It
overlaps well with the Monte Carlo spectrum generated at 9.46 GeV.

The $\eta'$ production rate is smaller at 9.46 GeV because of less
available energy. From the $q\bar{q}$ generator we found that the
production rate is 93.6\% that of 10.52 GeV. This factor is also
considered in estimation of the $\eta'$ production from $q\bar{q}$
events.

The mapping for continuum data is derived from the model-dependent
Monte Carlo spectrum. If the real data and the Monte Carlo are
very different then the systematic uncertainty due to this mapping
could be large. To check this, we compared the measured $\rm
E_{\eta'} / E_{beam}$ spectrum with the generated spectrum.
Fortunately, the spectra agree reasonably well and the systematic
uncertainty due to this source is negligible.

\begin{figure}[htb]
  \centerline{\epsfig{figure=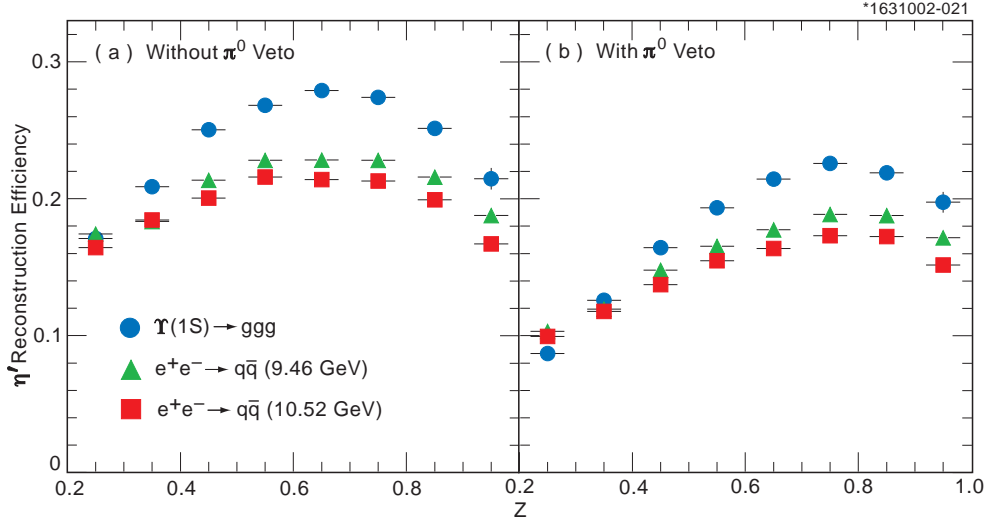,height=2.8in}}
\vspace{3mm}
\caption{\label{fig:eff}The $\eta'$ reconstruction efficiencies as function
 of Z for different MC samples a) without a $\pi^o$ veto, and
 b) with a $\pi^o$ veto in the photon selection.}
\end{figure}

We now turn to estimating the detection efficiencies. Shown in
Fig.~\ref{fig:eff} are the efficiencies estimated with different
models and different energies for a) without the $\pi^\circ$ veto,
and b) with the $\pi^\circ$ veto. In the real data we applied
$\pi^\circ$ veto to $\eta'$ candidates with Z $<$ 0.5. Comparing
with the efficiency from 9.46 GeV $q\bar{q}$ events, the
efficiency from $ggg$ events is roughly 15\% higher, and the
efficiency from 10.52 GeV $q\bar{q}$ events is roughly 7\% lower.
The main source of such difference is the event shape. The $ggg$
events are more spherical while the higher energy $q\bar{q}$ events
are more jetty.

\section{Extraction of the $\eta'$ Spectrum from $\Upsilon (1S)$ Decays}

The $\Upsilon (1S)$ data sample can be broken down into three
parts as described in the previous section:
$$N_{all} = N_{\Upsilon (1S)\to ggg} + N_{\Upsilon (1S)\to q\bar{q}} + N_{e^+e^-\to q\bar{q}}\ .$$
The first one has different reconstruction efficiencies from the other two.
For the contribution from continuum $(e^+e^-\to q\bar{q}$) events, we multiply the
number from off-resonance events at 10.52 GeV, mapped using Equation~\ref{eq:Z}, by a factor $f_{e^+e^-\to q\bar{q}}$
defined as:
\begin{equation}
N(\eta')_{e^+e^-\to q\bar{q}}(9.46\ {\rm GeV})=
N(\eta')_{e^+e^-\to q\bar{q}}(10.52\ {\rm GeV})\times
f_{e^+e^-\to q\bar{q}},
\end{equation}
where
\begin{eqnarray}
  f_{e^+e^-\to q\bar{q}}
   & = & \frac{80.4}{1193} \times \frac{1/9.46^2}{1/10.52^2} \times 0.9356
        \times \frac{\epsilon_{9.46}}{\epsilon_{10.52}} \nonumber \\
  & = & 0.078  \times \frac{\epsilon_{9.46}}{\epsilon_{10.52}}\ ,\label{eq:fgm}
\end{eqnarray}
where the first factor is the relative luminosities, the second the
energy squared dependence of the cross section, the third the
relative $\eta'$ yield and $\epsilon$ is the $Z$-dependent
reconstruction efficiency for $q\bar{q}$ events as shown
Fig.~\ref{fig:eff}.

We also want to evaluate the yield from $\Upsilon
(1S)\to\gamma^*\to q\overline{q}$. Since we know that
$\sigma_{\Upsilon (1S)\to\mu^+\mu^-} = 0.555\pm0.022\ {\rm nb}$
and $\sigma_{e^+e^-\to\mu^+\mu^-}(9.46\ {\rm GeV}) = 1.12\ {\rm
nb}$~\cite{Chen:1989ms}, we derive the factor to be
used in the $ N_{\Upsilon (1S)\to q\bar{q}}$ estimation as:
\begin{eqnarray}
  f_{\Upsilon (1S) \to q\bar{q}} & = & f_{e^+e^-\to q\bar{q}}
          \times \frac{R\cdot\sigma_{\Upsilon (1S)\to\mu^+\mu^-}}
          {R\cdot\sigma_{e^+e^-\to\mu^+\mu^-}} \nonumber \\
  & = & 0.0387  \times \frac{\epsilon_{9.46}}{\epsilon_{10.52}}. \label{eq:fqq}
\end{eqnarray}

In Table~\ref{tab:nrecon} we list the number of reconstructed
$\eta'$ over all $Z$ and in the high $Z$ region for various
$\Upsilon (1S)$ and continuum yields (only statistical errors are
shown). Note that the total numbers of signal from $\Upsilon (1S)$
data and off-resonance data in this table are the sum of all $Z$
bins derived bin per bin, as we need to use $Z$-dependent
efficiencies.

\begin{table}[ht]
\begin{center}
\begin{tabular}{lrr}
\hline \hline
 Sample                  &  All Z         & $\rm Z > 0.7$ \\ \hline
$\Upsilon (1S)$ data              & 1494 $\pm$ 120 &  46.0 $\pm$  8.1 \\
off-resonance            & 4294 $\pm$ 130 & 257.1 $\pm$ 17.3 \\
\hline
$\Upsilon (1S)\to ggg$            &  972 $\pm$ 120 &  13.9 $\pm$  8.1 \\
$\Upsilon (1S)\to q\bar{q}$       &  173 $\pm$   5 &  10.6 $\pm$  0.7 \\
Continuum $q\bar{q}$     &  349 $\pm$  11 &  21.5 $\pm$  1.4 \\
\hline
$\Upsilon (1S)\to ggg,q\bar{q}$   & 1145 $\pm$ 120 &  24.5 $\pm$  8.1 \\
\hline\hline
 \end{tabular}
\caption{\label{tab:nrecon} Number of reconstructed $\eta'$ from
$\Upsilon (1S)$ and off-resonance data
 and the breakdown categories of $\Upsilon (1S)$ data. Also listed are for samples with $\rm Z > 0.7$.}
\end{center}
\end{table}

The measured $\Upsilon (1S)\to\eta'X$ branching fractions are
listed in Table~\ref{tab:brresults} both for Z $>$ 0.7 and for all
Z. In the large Z region for 3 gluon decays, we do not have a
statistically significant signal and thus derive a 90\% confidence
level upper limit of ${\cal B}(\Upsilon (1S)\to ggg
\to\eta'X)_{Z>0.7}/{\cal B}(\Upsilon (1S)\to ggg)< 3.4\times
10^{-4}$. We describe the systematic errors below.

\begin{table}[ht]
\begin{center}
\begin{tabular}{lcc}
\hline \hline
Mode & All Z  &  Z $>$ 0.7\\\hline
  ${\cal B}(\Upsilon (1S)\to\eta'X)$ &
            $(2.8 \pm 0.4 \pm 0.2)$ \% & $(3.1 \pm 0.9 \pm 0.3)\times 10^{-4}$\\
  ${\cal B}(\Upsilon (1S)\to ggg\to\eta'X)/{\cal B}(\Upsilon (1S)\to ggg)$ &
            $(2.8 \pm 0.5 \pm 0.2) $\%&  $(1.9\pm 1.1 \pm 0.2)\times 10^{-4}$\\
 $ {\cal B}(\Upsilon (1S)\to q\bar{q}\to\eta'X)/{\cal B}(\Upsilon (1S)\to q\bar{q})$& $(4.2 \pm 0.2 \pm 0.4)$ \% & $(16.8\pm 1.1 \pm 1.7)
\times 10^{-4}$\\\hline\hline
\end{tabular}
\caption{\label{tab:brresults} Branching fractions of $\Upsilon
(1S)$ to $\eta'$ mesons, for all decays, three gluon decays and
quark-antiquark decays for the entire $\eta'$ energy spectrum and
for Z $>$ 0.7. The errors after the values give the statistical
and systematic uncertainties, respectively.}
\end{center}
\end{table}

The sources of systematic uncertainties are listed in
Table~\ref{tab:error} along with estimates of their sizes. The
total systematic errors on branching ratios are $\pm 10\%$ for
$q\bar{q}$ sample (independent of Z), $\pm 11\%$ for $ggg$ sample at $Z>0.7$, and
$\pm 8.6\%$ for the rest.

\begin{table}[ht]
\begin{center}
\begin{tabular}{lccc}
\hline \hline
 Sources     & $ggg$ ($Z>0.7$) & $q\bar{q}$ & All others \\ \hline
 Reconstruction efficiency of $\pi^\pm$  & 4.4 & 4.4 & 4.4 \\
 Reconstruction efficiency of $\eta$     &   5 &   5 &   5 \\
 Number of $\eta'$ from fit              &   2 &   2 &   2 \\
 Total number of $\Upsilon (1S)$                  & 2.4 & 2.4 & 2.4 \\
 ${\cal B}(\eta'\to\pi^+\pi^-\eta)$      & 3.4 & 3.4 & 3.4 \\ \hline
 ${\cal B}(\Upsilon (1S)\to q\bar{q})^{\dagger}$            &  -  & 3.2 & -   \\
 Ratio of integrated luminosity \cite{lumerr}         & 2.9 &  1  & -   \\
 $\sigma_{\Upsilon (1S)\to\mu^+\mu^-}$            & 3.6 &  4  & -   \\
 Z mapping                               &   6 &  3  & 3   \\ \hline
 Total                                   &  11 & 10  & 8.6 \\
\hline\hline
\multicolumn{4}{l}{$\dagger$ We use ${\cal B}(\Upsilon (1S)\to(q\bar{q}))=(8.83\pm 0.28)\%$.}
 \end{tabular}
\caption{\label{tab:error} The systematic uncertainties (in \%)
from different sources on the branching fraction measurements for
the 3 gluon sample for Z $>$ 0.7, the $q\bar{q}$ sample, and
both the 3 gluon sample for all Z and the total $\Upsilon (1S)$
sample.}
\end{center}
\end{table}

\begin{figure}[htb]
  \centerline{\epsfig{figure=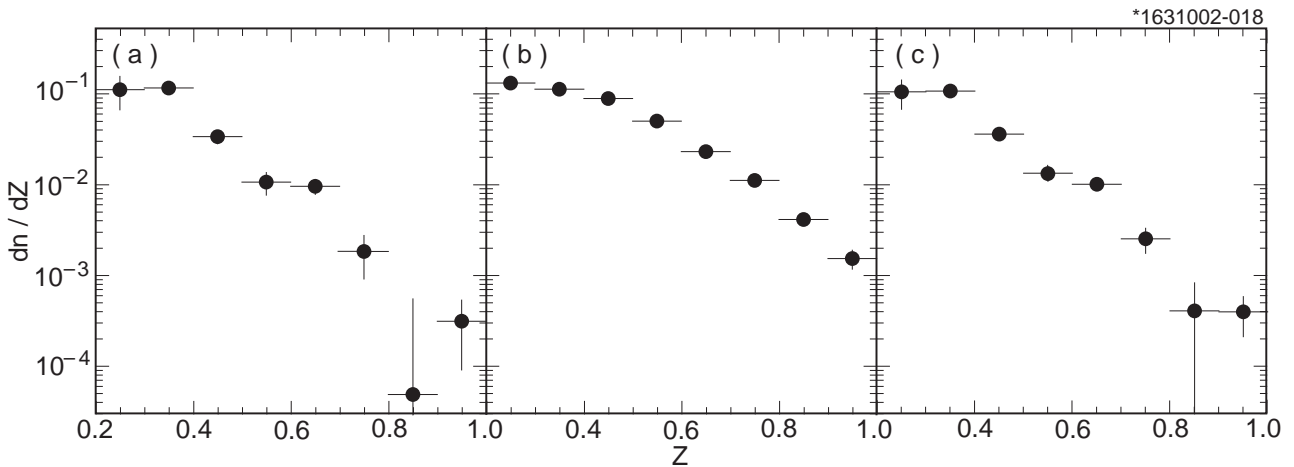,height=2.5in}}
\vspace{-3mm}
\caption{\label{fig:zspect}The differential branching fraction $dn/dZ$ as defined in
   context for a) $\Upsilon (1S)\to ggg\to\eta'X$, b) $\Upsilon (1S)\to q\bar{q}\to\eta'X$, and c) $\Upsilon (1S)\to\eta'X$.}
\end{figure}

We also measure the differential branching fractions as a function
of Z as shown in Fig.~\ref{fig:zspect}. In these plots only the
statistical error is shown, which dominates the total error.

 We define three relevant differential branching ratio's $dn/dZ$ as:
\begin{eqnarray}
  \frac{dn(ggg)}{dZ} &=& \frac{d {\cal B}(\Upsilon (1S)\to ggg\to\eta'X)}{dZ \times {\cal B}(\Upsilon (1S)\to ggg)}~~,  \nonumber \\
  \frac{dn(q \bar{q})}{dZ} &=& \frac{d {\cal B}(\Upsilon (1S)\to q \bar{q}\to\eta'X)}{dZ \times {\cal B}(\Upsilon (1S)\to q \bar{q})}~~,  \nonumber \\
  \frac{dn(1S)}{dZ} &=& \frac{d {\cal B}(\Upsilon (1S)\to\eta'X)}{dZ}~~.
\end{eqnarray}

Listed in Table~\ref{tab:branching} are the differential branching
fractions in Z intervals for $\Upsilon (1S)$ decays to $\eta'$ for
$ggg$ and $q\bar{q}$ subsamples and all decays.
\begin{table}[ht]
\begin{center}
\begin{tabular}{cccc}
\hline \hline
  Z        &  $\Upsilon (1S)\to(ggg)$ & $\Upsilon (1S)\to(q\bar{q})$ & All $\Upsilon (1S)$\\ \hline
 0.2 - 0.3 & 11164 $\pm$ 4471 & 13205 $\pm$ 1253 & 10503 $\pm$ 3740\\
 0.3 - 0.4 & 11624 $\pm$ 1314 & 11250 $\pm$  685 & 10716 $\pm$ 1099\\
 0.4 - 0.5 &  3381 $\pm$  558 &  8898 $\pm$  416 &  3614 $\pm$  467\\
 0.5 - 0.6 &  1067 $\pm$  300 &  5030 $\pm$  272 &  1336 $\pm$  251\\
 0.6 - 0.7 &   963 $\pm$  181 &  2321 $\pm$  166 &  1011 $\pm$  151\\
 0.7 - 0.8 &   184 $\pm$   92 &  1116 $\pm$  102 &   252 $\pm$   77\\
 0.8 - 0.9 &     5 $\pm$   50 &   415 $\pm$   59 &    41 $\pm$   42\\
 0.9 - 1.0 &    31 $\pm$   22 &   153 $\pm$   36 &    40 $\pm$   19\\ \hline
 0.7 - 1.0 &    19 $\pm$   11 &   168 $\pm$   11 &    31 $\pm$    9\\\hline
sum of all &  2842 $\pm$  471 &  4239 $\pm$  153 &  2751 $\pm$  394\\

\hline\hline
 \end{tabular}
\caption{\label{tab:branching} Differential branching fractions of
$\eta'$ ($\times 10^{-5}$). The last two rows are total branching
fractions. The branching fractions in columns 2 and 3 are
normalized to the total branching fraction of $\Upsilon
(1S)\to(ggg)$ and $\Upsilon (1S)\to(q\bar{q})$ respectively, while
the last column is normalized to all $\Upsilon (1S)$ decay. The errors are
statistical only, the systematic errors on the absolute normalization
for column 1 is 8.6\% for Z $<$ 0.7, 11\% for Z $>$ 0.7, and 10\% and 8.6\%
for columns 2 and 3, respectively.}
\end{center}
\end{table}

In the Z spectrum of $\eta'$ mesons produced via $ggg$, there is an
excess above an apparent exponential decrease for 0.6 $<$ Z $<$ 0.7,
corresponding to a recoil mass opposite the $\eta'$ in the range
5.3 to 6.1 GeV \cite{excess}. However, a detailed study did not
reveal any narrow structures. A possible explanation is that there
is more than one process contributing to this distribution. We
note also that the $q\bar{q}$ has much larger rates at high Z than
$ggg$.

\section{Comparison with Theory and Conclusions}

\begin{figure}[htb]
  \centerline{\epsfig{figure=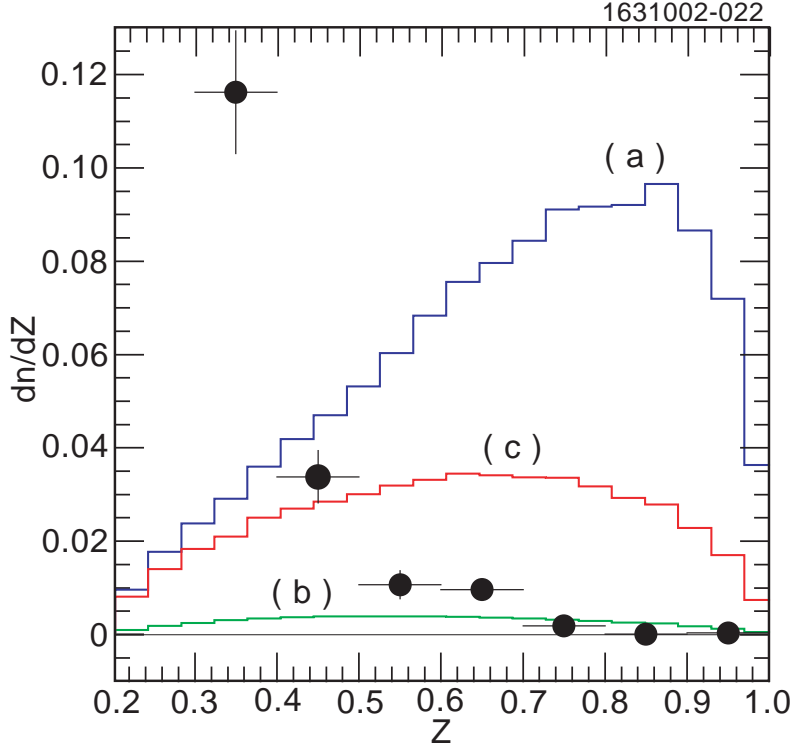,height=4in}}
\vspace{-5mm}
\caption{\label{fig:spec}
   The measured $dn/dZ$ spectrum of $\Upsilon (1S)\to (ggg)\to\eta'X$ compared with
   theoretical predictions. Shown in dots are the measurement in this study.
   Shown in lines are different theoretical predictions: a) a slowly falling
   form factor, b) a rapidly falling form factor, and c) intermediate form
   factor~\cite{Kagan02}. These predictions are valid only in the region Z $>$
   0.7.}
\end{figure}

Fig.~\ref{fig:spec} shows the Z spectrum of the $\eta'$ measured in
this paper compared with the spectra predicted by the three
different models described above. The models are expected to dominate $\eta'$ production only for Z$> 0.7$, with other fragmentation based processes being important at lower Z. The measurement strongly favors a
rapidly falling $q^2$ dependence of the ${g^*g\eta'}$ form factor
predicted by pQCD \cite{Ali:2000ci,Muta}, and ruling out other models.

In conclusion, we have made the first measurement of the $\eta'$
energy spectrum from $\Upsilon (1S)\to ggg$ decays. Our data are not
consistent with an enhanced $\eta'g^*g$ coupling at large $\eta'$
energies. Thus, the large observed $\eta'$ yield near end point of
the charmless $B$ decay spectrum cannot be explained by a large
$\eta'g^*g$ form-factor. Therefore, new physics has not been ruled
out and may indeed be present in rare $b$ decays.

\section{Acknowledgments}
We thank Alex Kagan for providing us with his calculations and we thank A. Kagan and A. Ali for useful discussions on the theoretical models.
We gratefully acknowledge the effort of the CESR staff in providing us with
excellent luminosity and running conditions.
M. Selen thanks the Research Corporation,
and A.H. Mahmood thanks the Texas Advanced Research Program.
This work was supported by the National Science Foundation and the
U.S. Department of Energy.

\clearpage


\begin{thebibliography}{99}
\bibitem{lumerr}
We use the ratio of integrated luminosities of $\Upsilon (1S)$ and
off-resonance data in calculating $f_{e^+e^-\to q\bar{q}}$ and
$f_{\Upsilon (1S) \to q\bar{q}}$. The uncertainty of this ratio is
about $\pm 1\%$. The uncertainty of $\sigma_{\Upsilon
(1S)\to\mu^+\mu^-}$, 4\%, also affects $f_{\Upsilon (1S) \to
q\bar{q}}$. These latter two directly affect the branching
fraction in the $q\bar{q}$ sample with uncertainties of 1\% and
4\% respectively. The effects of these two sources to the overall
and the $ggg$ sample branching fractions are negligible except for
the branching fraction measurement of high energy $\eta'$ in the
$ggg$ samples, where there are $\pm 2.9\%$ and $\pm 3.6\%$
uncertainties.
\bibitem{excess}
We fit the Z distribution in the range 0.3 $<$ Z $<$ 1, not
including the point 0.6 $<$ Z $<$ 0.7, to an assumed inherent
exponential shape and determined that this point is 3.3 standard
deviations in excess.
\end{thebibliography}
\end{document}